\documentstyle[preprint,prl,aps]{revtex}
\begin{document}
\draft

\title{Separability criterion and inseparable mixed states with \\
positive partial transposition }

\author{Pawe\l{} Horodecki}

\address{Faculty of Applied Physics and Mathematics\\
Technical University of Gda\'nsk, 80--952 Gda\'nsk, Poland}

\maketitle

\begin{abstract}
It is shown that any separable state on Hilbert space
${\cal H}={\cal H}_1\otimes{\cal H}_2$, can be written as a
convex combination of N pure product states with $N\leq 
(dim{\cal H})^2$.  Then a new separability criterion for 
mixed states in terms of range of density matrix is obtained.
It is used in construction of inseparable mixed states with 
{\it positive} partial transposition in the case of $3\times 3$ 
and $2\times 4$ systems.  The states represent an entanglement 
which is hidden in a more subtle way than it has been known so far.
\end{abstract}

\newtheorem{theorem}{Theorem}
\newtheorem{lemma}{Lemma}
\section{Introduction}
The problem of quantum inseparability
of mixed states has attracted much attention recently and it
has been widely considered  in different physical contexts (see \cite{duza}
and references therein).
In particular effective criterion of separability of $2\times 2$ and
$2\times 3$ systems has been obtained  \cite{Peres,my}.
Quite recently the criterion has been used for characterisation of
two-bit quantum gate \cite{Cirac} and quantum broadcasting
\cite{buzek}. It also allowed to show that
any inseparable state of $2 \times 2$ system can be distilled to
a singlet form \cite{dest}.

Recall that the state $\varrho$
acting on the Hilbert space ${\cal H}={\cal H}_1 \otimes {\cal H}_2$
is called separable
if it can be wiritten or approximated (in the trace norm) by the states of the form
\begin{equation}
\varrho=\sum_{i=1}^kp_i\varrho_i\otimes\tilde \varrho_i
\label{sep}
\end{equation}
where $\varrho_i$ and $\tilde\varrho_i$ are states on ${\cal H}_1$ and
${\cal H}_2$ respectively.
Usually one deals with the case $dim{\cal H}=m$.
For this case it will appear subsequently that
any separable state can be written as a convex combination of
finite product pure states i.e. in those cases the
``approximation'' part of the definition appears to be redundant.

Peres has shown \cite{Peres} that the necessary condition for separability
of the state $\varrho$ is positivity of its
partial transposition $\varrho^{T_2}$.
The latter
associated with an arbitrary product orthonormal $f_i \otimes f_j$
basis is defined by the matrix elements:
\begin{equation}
\varrho^{T_2}_{m\mu,n\nu}\equiv
\langle f_m\otimes f_\mu| \varrho^{T_2}| f_n \otimes f_\nu\rangle=
\varrho_{m\nu,n\mu}.
\label{tr}
\end{equation}
Although the matrix $\varrho^{T_2}$ depends on
the used basis, its eigenvalues do not.
Consequently, for any state the condition can be checked
using {\it an arbitrary} product orthonormal basis.
\footnote{As the full transposition of positive operator is also
positive, the positivity of the partial transposition  $\varrho^{T_2}$
is equivalent to positivity of (defined in analogous way)
partial transposition  $\varrho^{T_1}$.}
It has been shown \cite{my} that for the systems $2 \times 2$
and $2 \times 3$
the partial transposition condition is also sufficient one.
Thus in those cases the set of separable states
has been characterized completely in a simple way.

For higher dimensions
the necessary and sufficient condition for separability
has been provided \cite{my} in terms of positive maps.
Namely the state $\varrho$ acting on Hilbert space
${\cal H}={\cal H}_1 \otimes {\cal H}_2$ is separable iff for any positive map
$\Lambda : {\cal B}({\cal H}_2) \rightarrow {\cal B}({\cal H}_1)$
the operator $I\otimes\Lambda\varrho$ is positive
(${\cal B}({\cal H}_i)$ denote the set of all operators
acting on ${\cal H}_i$ and I is identity map).
Then a natural question arose, whether the partial transposition
condition is also sufficient for higher dimensions.
The negative answer to this question
has been established \cite{my} with no,
however, explicit counterexample given.
In this Letter we provide (Sec. 3) a new criterion for separability
of quantum states.
It is done with the help of analysis of
range of density matrices via result on
the decomposition of separable states on
pure product states (Sec. 2). Namely it appears that
any separable states can be written as a convex
combination of {\it finite} number  N of pure product states
with N restricted by squared dimension of
respective Hilbert space.
In Sec. 4 we construct families of inseparable states with
{\it positive} partial transposition for $3\times 3$ and $2\times 4$ systems.
We achieve our goal using the separability criterion and
the technique introduced by Woronowicz in his paper (1976)
\cite{Woronowicz}
which has provided heuristic basis for the present analysis.
It appears that, in general, the new criterion is
rather independent than equivalent to the partial transposition
one.

\section{Finite decomposition of separable states}
We prove here the following
\begin{theorem}
Let $\varrho$ is a separable state acting on the
Hilbert space ${\cal H}={\cal H}_1\otimes{\cal H}_2$,
$dim {\cal H}=m< \infty$.
Then there exists set of N, $N\leq m^2$
product projectors $P_{\psi_i}\otimes Q_{\phi_k}$,
\{i,k\} $\in$ I
(I is a finite set of pairs of indices with number of
pairs $N=\#I\leq m^2$) and probabilities $p_{ik}$
such that
\begin{equation}
\varrho=\sum_{\{i,k\} \in I}p_{ik}P_{\psi_i}\otimes Q_{\phi_k}.
\end{equation}
\end{theorem}
{\it Proof.-} The proof depends on properties of
compact convex sets in real finitedimensional spaces.
Namely then the set of separable states $M_{sep}$ defined above,
can be treated as a compact convex subset of finite-dimensional real space
(obtained by real linear combinations 
of Hermitian operators) with a dimension $n=m^2-1$.
The set of separable states $M_{sep}$
is a closed convex subset of covex set of all quantum
states $P$ which is bounded, compact and given by
$n=m^2-1$ real parameters in Hilbert-Schmidt basis.

Let us denote by $P_{sep}\subset M_{sep}$ set of all
separable pure states. $P_{sep}$ is obviously compact
(it is a tensor product of two spheres which are
compact in finite dimensional case).
We have from the definition of $M_{sep}$ immedately :
\begin{equation}
M_{sep}=\overline{ conv P_{sep}}.
\end{equation}
Here conv A denotes convex hull of A and means
set of all possible finite convex combinations
of elements from A \footnote{The convex hull of the set of A is
usually defined as a minimal convex set containing A,
but it is shown that
it is equivalent to the set of all possible finite
convex combinations of element of A (see Ref.\cite{Kelly}).}.
$\overline{B}$ stands for the closure of B in the
trace norm topology.
It appears as the standard fact from the
convex set theory that convex hull
of any compact set from finite dimensional
space is compact itself (see Ref. \cite{Kelly}, theorem 14, p. 210).
Thus closure of $M_{sep}=\overline{conv P_{sep}}=conv P_{sep}$.
Hence set of extremal points of $M_{sep}$ if equivalent
to $P_{sep}$. Then we can apply
the Caratheodory theorem \cite{Alfsen}
which says that any element of
compact convex subset of ${\cal R}^n$ (in our case
$n=m^2-1$)
can be represented as a convex combination of
(at most n+1) afinely independent extreme points \cite{Alfsen}.
Usage of this theorem completes the proof of our statement.
\section{Separability criterion}
First we will need the following
\begin{lemma}
Let state $\varrho$ act on the
Hilbert space ${\cal H}$, $dim {\cal H}< \infty$ .
Then for an arbitrary $\varrho$--ensemble $\{ \Psi_i, p_i \}$:
\begin{equation}
\varrho=\sum_{i}p_i|\Psi_i><\Psi_i|
\end{equation}
each of vectors $\Psi_i$ belongs to the range of the state $\varrho$.
\end{lemma}
{\it Proof.-}
The range of $\varrho$ is defined by
$Ran\varrho\equiv \{ \psi \in {\cal H}$:
$\varrho \phi=\psi$ for some $\phi \in {\cal H} \}$.
As $\varrho$ is linear and hermitian operator
we have that
$Ran\varrho$ is simply a subspace spanned
by all eigenvectors of $\varrho$ belonging
to nonzero eigenvalues. In short, $Ran\varrho$
is a support of $\varrho$.
Following \cite{Wooters} we have that $\Psi_i$ belongs to the support of
$\varrho$. Thus any $\Psi_i$ belongs to $Ran\varrho$.

Now we can prove our main result:
\begin{theorem}
Let $\varrho$ act on Hilbert space ${\cal H}={\cal H}_1 \otimes {\cal H}_2$, 
$dim{\cal H}=m$ .
If $\varrho$ is separable then there exists a set of
product vectors \{$\psi_i\otimes\phi_k$\}, \{i,k\} $\in$ I
(I is a finite set of pairs of indices with number of
pairs $N=\#I\leq m^2$) and
probabilities $p_{ik}$ such that

(i)  the ensemble   \{$\psi_i\otimes\phi_k$, $p_{ik}$ \} ,
(   \{$\psi_i\otimes\phi_k^{\star}$, $p_{ik}$ \} ) corresponds to
the matrix $\varrho$, ($\varrho^{T_2}$),

(ii) the vectors \{$\psi_i\otimes\phi_k$ \},
(\{$\psi_i\otimes\phi_k^{\star}$ \}) span
the range of $\varrho$ ($\varrho^{T_2}$), in particular
any of vectors \{$\psi_i\otimes\phi_k$\}
(\{$\psi_i\otimes\phi_{k}^{\star}$\}) belongs to the range of
$\varrho$ ($\varrho^{T_2}$).
\label{dod}
\end{theorem}
{\it Proof.-}
Let us prove first the statement (i).
According to the theorem form Section II
any separable state $\varrho$ can be written
in the form
\begin{equation}
\varrho=\sum_{\{i,k\} \in I}p_{ik}P_{\psi_i}\otimes Q_{\phi_k}\equiv
\sum_{\{i,k\} \in I}p_{ik}|\psi_i\otimes \phi_k><\psi_i\otimes \phi_k|.
\end{equation}
using only $N=\#I\leq m^2$
pure product states $P_{\psi_i}\otimes Q_{\phi_k}$.
Remembering that the transposition of Hermitian operator is simply
equivalent to the complex conjugation of its matrix elements
we get
\begin{equation}
Q_{\phi_k}^T=Q_{\phi_k}^{\star}= (|\phi_k><\phi_k|)^{\star}=
|\phi_k^{\star}><\phi_k^{\star}|= Q_{\phi_k^{\star}}.
\end{equation}
From the above and the definition of partial transposition
(\ref{tr}) we obtain
\begin{equation}
\varrho^{T_2}=\sum_{\{i,k\} \in I}p_{ik}P_{\psi_i}\otimes Q_{\phi_k}^T\equiv
\sum_{\{i,k\} \in I}p_{ik}|\psi_i\otimes \phi_k^{\star}><\psi_i\otimes \phi_k^{\star}|.
\end{equation}
hence we obtain the statement (i).
Obviously, any vector $\psi$ from the range of the state
is given by a linear combination of vectors belonging to
the ensemble realising the state.
Using the lemma immediately completes the proof of (ii).

{\it Remark 1.-} Using the full transposition one can easily see
that the analogous theorem (with vectors conjugated
on the first space) equivalent to the above one
is valid for $\varrho^{T_1}$.

{\it Remark 2.-}
The conjugation $\phi^{\star}$ associated with the basis the transposition of
$Q_{\phi}$ was performed in is simply obtained by complex conjugation
of the coefficients in this basis up to the irrelevant phase
factor. Then the operation of partial complex conjugation (we will
denote it by $\Psi^{\star 2}$ ) can be illustrated as follows
\begin{equation}
((\alpha e_1 + \beta e_2) \otimes (\gamma e_1 + \delta e_2))^{\star 2}\equiv
(\alpha e_1 + \beta e_2) \otimes (\gamma e_1 + \delta e_2)^{\star} \equiv
(\alpha e_1 + \beta e_2) \otimes (\gamma^{\star} e_1 + \delta^{\star} e_2),
\end{equation}
where the standard basis $e_1$, $e_2$ in ${\cal C}^2$
was used in the transposition of corresponding projector.
Note that the operation of partial conjugation is defined {\it only}
for product vectors.
\section{Inseparable states with positive partial transposition}
{\it A. $3\times 3$ system .-} Consider the Hilbert space
${\cal H}={\cal C}^3 \otimes {\cal C}^3$.
Let  $P_{\phi}\equiv |\phi><\phi|$
and let $\{ e_i \}$, i=1, 2, 3 stand for standard basis
in ${\cal C}^3$. Then we define projector
\begin{equation}
Q\equiv I\otimes I - (\sum_{i=1}^{3}P_{e_i}\otimes P_{e_i} + P_{e_3}\otimes P_{e_1})
\end{equation}
and vectors
\begin{equation}
\Psi\equiv {1 \over \sqrt{3}}(e_1\otimes e_1+e_2\otimes e_2+e_3\otimes e_3),
\label{ent}
\end{equation}
\begin{equation}
\Phi_{a}\equiv e_3\otimes(\sqrt{{1+a \over 2}}e_1 +
\sqrt{{1-a \over 2}}e_3), \\\\\\ 0\leq a\leq 1
\label{prod}
\end{equation}
Now we define the following state
\begin{equation}
\varrho_{insep}\equiv {3 \over 8}P_{\Psi} + {1 \over 8}Q.
\end{equation}
This state is inseparable as its partial transposition
possesses a negative eigenvalue $\lambda={1-\sqrt{5} \over 2}$
belonging to the eigenvector ${2 \over 5+\sqrt{5}}(e_1\otimes e_3 +
{-1-\sqrt{5} \over 2}e_3\otimes e_1)$. Here inseparability comes
from highly entangled pure state $P_{\Psi}$.
On the other hand the state $P_{\Phi_a}$ corresponding
to the vector (\ref{prod}) is evidently separable.
Below we will see that it is possible to mix the states $\varrho_{insep}$
and $P_{\Phi_a}$ in such a way that the resulting state will have
partial transposition positive being nevertheless inseparable.
For this purpose consider the following state
\begin{equation}
\varrho_a={8a \over 8a + 1}\varrho_{insep} + {1 \over 8a + 1}P_{\Psi_a}.
\label{state}
\end{equation}
Its matrix and the matrix of its partial transposition are of the form
\begin{eqnarray}
\varrho_a={1 \over 8a + 1}
\left[ \begin{array}{ccccccccc}
          a &0&0&0&a&0&0&0& a   \\
           0&a&0&0&0&0&0&0&0     \\
           0&0&a&0&0&0&0&0&0     \\
           0&0&0&a&0&0&0&0&0     \\
          a &0&0&0&a&0&0&0& a     \\
           0&0&0&0&0&a&0&0&0     \\
           0&0&0&0&0&0&{1+a \over 2}&0&{\sqrt{1-a^2} \over 2}\\
           0&0&0&0&0&0&0&a&0     \\
          a &0&0&0&a&0&{\sqrt{1-a^2} \over 2}&0&{1+a \over 2}\\
       \end{array}
      \right ], \ \ \
\varrho_a^{T_2}=
{1 \over 8a + 1}\left[ \begin{array}{ccccccccc}
          a &0&0&0&0&0&0&0&0   \\
           0&a&0&a&0&0&0&0&0     \\
           0&0&a&0&0&0&a&0&0     \\
           0&a&0&a&0&0&0&0&0     \\
           0&0&0&0&a&0&0&0&0     \\
           0&0&0&0&0&a&0&a&0     \\
           0&0&a&0&0&0&{1+a \over 2}&0&{\sqrt{1-a^2} \over 2}\\
           0&0&0&0&0&a&0&a&0     \\
           0&0&0&0&0&0&{\sqrt{1-a^2} \over 2}&0&{1+a \over 2}\\
       \end{array}
      \right ]. \nonumber
\end{eqnarray}

It is easy to show that $\varrho_a^{T_2}$ is positive. Indeed
it suffices only to single out the state $I\otimes U P_{\Phi_a}I\otimes U^{\dagger}$
as a component of convex combination
where
\begin{equation}
U=\left[ \begin{array}{ccc}
           0&0&1 \\
           0&1&0 \\
           1&0&0 \\
       \end{array}
      \right ]
\label{matrix}
\end{equation}
and then check that the remaining operator in the combination
is positive. Thus $\varrho_a^{T_2}$ is a legitimate state.
Now we will show that it is inseparable. Then, as the operation of partial
transposition preserves separability, we will have two
``dual'' sets of inseparable mixtures with positive partial transposition.
Let us find all product (unnormalised for convenience) vectors
belonging to the range of $\varrho_a^{T_2}$.
We will adopt here the horizontal notation with
basis ordered in the following way
$e_1\otimes e_1$, $e_1\otimes e_2$, $e_1\otimes e_3$,
$e_2\otimes e_1$, $e_2\otimes e_2$... and so on.
Assume, in addition, that $a\neq 0, 1$. Then any vector belonging to the
range of $\varrho_a^{T_2}$ can be presented as
\begin{equation}
u=(A,B,C;B,D,E;C+F,E,xF),\\\ A,B,C,D,E,F \in {\cal C},
\label{ob}
\end{equation}
with nonzero
$x=\sqrt{1+a \over 1-a}$.
On the other hand if $u$ is to be positive, it must be of the form
\begin{equation}
u_{prod}=(r,s,t)\otimes (\tilde A,\tilde B,\tilde C)
\equiv(r(\tilde A,\tilde B,\tilde C);s(\tilde A,\tilde B,\tilde C);
t(\tilde A,\tilde B,\tilde C)), \\\
r,s,t, \tilde A, \tilde B,\tilde  C \in {\cal C}.
\label{il}
\end{equation}
Let us now consider the following cases:

i) $rs\neq 0$, then without loss generality we can
put $r=1$ and $\tilde A=A$, $\tilde B=B$, $\tilde C=C$.
Comparison with (\ref{ob}) gives us in turn: $B=sA$; $E=sC$, $E=tsA$ $\Rightarrow$
$C=tA$ (hence $A\neq 0$ or $u_{prod}$ vanishes); $xF=tC=t^2A$, $C+F=tA$ with $C=tA$
$\Rightarrow$
$F=0$; $xF=tC=t^2A$ in the presence of vanishing $F$ and non vanishing $A$
$\Rightarrow$ $t=0$.
Thus we obtain the states
\begin{equation}
u_1=A(1,s,0)\otimes(1,s,0), \\\ A, s \in {\cal C}.
\label{u1}
\end{equation}

ii) $r=0$. Then we have
\begin{equation}
u_{prod}=(0,0,0;s(\tilde A,\tilde B, \tilde C);t(\tilde A,\tilde B,\tilde C)), \\\
s,t, \tilde A, \tilde B, \tilde C \in {\cal C} .
\label{pro1}
\end{equation}
On the other hand one gets
\begin{equation}
u_{prod}=(0,0,0;0,D,E;F,E,xF), \\\ D,E,F \in {\cal C}.
\label{prod1}
\end{equation}
Now either $s=0$ and then, according to (\ref{prod1})
$E=0$ gives us
\begin{equation}
u_2=F(0,0,1)\otimes (1,0,x), \\\ F \in {\cal C}
\label{u2}
\end{equation}
or $s\neq 0$.
In the last case we can put $s=1$.
Consequently it is possible that $t=0$
and then we get via conditions $F=0$, $E=0$ another product state
\begin{equation}
u_3=D(0,1,0)\otimes (0,1,0) , \\\ D \in {\cal C}.
\label{u3}
\end{equation}
For the case $t\neq 0$
we get from (\ref{pro1}), (\ref{prod1}) $\tilde A=0$ $\Rightarrow$
$F=0$ $\Rightarrow$  $E=0$ $\Rightarrow$ $D=0$ . Hence the only
product vector with non vanishing $t$ is trivial zero vector.

iii) $r\neq 0$, $s=0$. As in the case of (i) we can put
$r=1$ and $\tilde A=A$, $\tilde B=B$, $\tilde C=C$.
Then we have $B=E=D=0$
which leads to the equality
\begin{equation}
(A,0,C;0,0,0;t(A,0,C))=(A,0,C;0,0,0;C+F,0,xF).
\label{pro}
\end{equation}
Then for $t=0$ we get $C=F=0$
and
\begin{equation}
u_4=A(1,0,0)\otimes (1,0,0) , \\\ A \in {\cal C},
\label{u4}
\end{equation}
or, provided that $t\neq 0$ , $xF=tC$ , $C+F=tA$ $\Rightarrow$
$A=(t^{-1}+x^{-1})C$ and then
\begin{equation}
u_5=C(1,0,t)\otimes (t^{-1}+x^{-1} ,0,1 ), \\\  C, t \in {\cal C}, t\neq 0.
\label{u5}
\end{equation}
All partial complex conjugations of vectors
(\ref{u1}), (\ref{u2}), (\ref{u3}), (\ref{u4}), (\ref{u5}) are
\begin{eqnarray}
&&u_1^{\star 2}=A(1,s,0)\otimes(1,s^{\star},0), A, s \in {\cal C} , s\neq 0,
 \nonumber \\
&&u_2^{\star 2}=F(0,0,1)\otimes (1,0,x) , F \in {\cal C}, \nonumber \\
&&u_3^{\star 2}=D(0,1,0)\otimes (0,1,0) , D \in {\cal C}, \nonumber \\
&&u_4^{\star 2}=A(1,0,0)\otimes (1,0,0) , A \in {\cal C}, \nonumber \\
&&u_5^{\star 2}=C(1,0,t)\otimes ((t^{\star})^{-1}+x^{-1} ,0,1) ,
C, t  \in {\cal C}, t \neq 0.
\end{eqnarray}
It is easy to see that the above vectors can not span the range of $\varrho_a$
as they are orthogonal to the vector
\begin{equation}
\tilde u=(0,0,1)\otimes(0,1,0).
\end{equation}
which belongs just to $Ran\varrho_a$.
Hence, for any $a\neq 0, 1$ ,
the state $\varrho_a^{T_2}$ violates the condition
due to the second statement of the theorem. Thus
the state is inseparable together with the ``dual'',
original state $\varrho_a$.
In the latter the state $P_{\Phi_a}$ masks the inseparability due to
$\varrho_{insep}$, making it ``invisible''
to the partial transposition criterion, but does not destroy it.

It is interesting to see the limit behaviour of the
state $\varrho_a$.
In the case of $a=0$ we get the separable state with
the symmetric representation
\begin{equation}
\varrho_0\equiv{3 \over 9}P_{\Psi}+{1 \over 9}(I\otimes I-
\sum_{i=1}^{3}P_{e_i}\otimes P_{e_i})={1 \over 9}\int_0^{2\pi}
P_{\Phi(\phi)}\otimes P_{\Phi(\phi)}{d\phi \over 2\pi},
\label{int}
\end{equation}
where projectors $P_{\Phi(\phi)}$
correspond to the vectors
$\Phi(\phi)=1/\sqrt{3}(1,e^{i\phi},e^{-2i\phi})$.

Note that the integral  representation (\ref{int}) is not unique.
The representation of the ``dual'' state $\varrho_a^{T_2}$
is obtained by complex conjugation of projectors acting
on the second space.
For the case $a=0$ we get simply the product
state $P_{\Phi_1}$ (c.f. (\ref{prod})).
Taking the
parameter $a$ arbitrarily
close to $0$, we obtain almost product pure states $P_{\Phi_a}$
being nevertheless separable.
The situation is, in a sense, analogical to the case of the states introduced in
\cite{tata}. The inseparability of the latter was also
determined by parametric change of both coherences
and probabilities involved in the state.

{\it B. $2\times 4$ system .-} Here we will use the vectors
\begin{equation}
\Psi_i={1 \over \sqrt{2}}(e_1\otimes e_i+e_2\otimes e_{i+1}), \ \ i=1, 2, 3,
\end{equation}
\begin{equation}
\Phi_{b}\equiv e_2\otimes(\sqrt{{1+b \over 2}}e_1 +
\sqrt{{1-b \over 2}}e_3), \\\ 0\leq b\leq 1.
\label{pr}
\end{equation}
Then we can construct the following state
\begin{equation}
\sigma_{insep}={2 \over 7}\sum^{3}_{i=1}P_{\Psi_i} + {1 \over 7}P_{e_1\otimes e_4},
\end{equation}
which is inseparable (it can be easily verified like in the state
$\sigma_{insep}$ using partial transposition criterion).
Now the states of our interest are of the form
\begin{equation}
\sigma_b={7b \over 7b + 1}\sigma_{insep} + {1 \over 7b + 1}P_{\Phi_b}.
\end{equation}
The corresponding matrices are \footnote{The example of pair
of matrices of such a type treated, however, as operators on
${\cal C}^{4}  \oplus {\cal C}^{4}$ together with similar analysis of
their ranges has been considered in \cite{Woronowicz} in the context
of positive maps.}
\begin{eqnarray}
\sigma_b={1 \over 7b + 1}
\left[ \begin{array}{cccccccc}
           b&0&0&0&0&b&0&0   \\
           0&b&0&0&0&0&b&0     \\
           0&0&b&0&0&0&0&b     \\
           0&0&0&b&0&0&0&0     \\
           0&0&0&0&{1+b \over 2}&0&0&{\sqrt{1-b^2} \over 2} \\
           b&0&0&0&0&b&0&0     \\
           0&b&0&0&0&0&b&0     \\
           0&0&b&0&{\sqrt{1-b^2} \over 2}&0&0&{1+b \over 2}\\
       \end{array}
      \right ], \ \ \
\sigma_b^{T_2}={1 \over 7b + 1}
\left[ \begin{array}{cccccccc}
           b&0&0&0&0&0&0&0   \\
           0&b&0&0&b&0&0&0     \\
           0&0&b&0&0&b&0&0     \\
           0&0&0&b&0&0&b&0     \\
           0&b&0&0&{1+b \over 2}&0&0&{\sqrt{1-b^2} \over 2}\\
           0&0&b&0&0&b&0&0     \\
           0&0&0&b&0&0&b&0     \\
           0&0&0&0&{\sqrt{1-b^2} \over 2}&0&0&{1+b \over 2}\\
       \end{array}
      \right ]
\end{eqnarray}

It is easy to see that the state $\sigma_{b}^{T_2}$
is positive as
\begin{equation}
\sigma_{b}^{T_2}=I\otimes U\sigma_{b} I\otimes U^{\dagger},
\end{equation}
with
\begin{equation}
U=\left[ \begin{array}{cccc}
           0&0&0&1 \\
           0&0&1&0 \\
           0&1&0&0 \\
           1&0&0&0 \\
       \end{array}
      \right ].
\label{matr}
\end{equation}

Consequently, for
$y=\sqrt{1-b \over 1+b}\neq 0,1$ we get,
analogically as in the part A of the section,
the partial complex conjugations of
all possible product vectors $v_i \in$ $Ran\sigma_{b}$
\begin{eqnarray}
&&v_1^{\star 2}=C(1,s)\otimes((s^{\star})^2,s^{\star},1,(s^{\star})^{-1}
(1+y(s^{\star})^3)), \\\
C, s \in {\cal C} , s\neq 0, \nonumber \\
&&v_2^{\star 2}=F(0,1)\otimes (1,0,0,y), \ F \in {\cal C} \nonumber \\
&&v_3^{\star 2}=D(1,0)\otimes (0,0,0,1) , \ D \in {\cal C}.
\label{sprz}
\end{eqnarray}
On the other hand, any vectors from
the range of $\sigma_{a}^{T_2}$ can be written in our notation as
\begin{equation}
w=(A', B', C' , D' ;B' + yE', C', D', E'), \\\\\   A',B',C',D',E' \in {\cal C}.
\label{prim}
\end{equation}
Let us check now whether the vectors (\ref{sprz})
can be written in the above form.
For the  $v_1^{\star 2}$, assuming that it is nontrivial one ($C\neq 0$)
and at the same time is of the from (\ref{prim}),
taking into account the coefficient
$C'$ we obtain
\begin{equation}
s^{\star}=s^{-1}.
\end{equation}
On the other hand, considering $B'$, $E'$ and $B'+ yE'$, we have
\begin{equation}
s^{\star}+ys(s^{\star})^{-1}(1+y(s^{\star})^3)=s(s^{\star})^{2}.
\end{equation}
Finally taking into account $D'$ we obtain
\begin{equation}
(s^{\star})^{-1}(1+y(s^{\star})^3)=s.
\end{equation}
Combining all the three equations above we find that $ys^2=0$, $s \neq 0$ which
contradicts the fact that $y$ does not vanish.
For $v_2^{\star 2}$ we obtain $yE'=D$ and at the same
time $E'=yD$, which is impossible for $y\neq 0, 1$ unless
$E'=D=0$ trivialising then the vector $v_2^{\star 2}$. For the
vector $v_3^{\star 2}$ we get immediately
that it must hold $D=0$. It leads to the conclusion
that {\it none} of vectors $v_i^{\star 2}$ belongs to the
$Ran\sigma_b^{T_2}$ apart from the trivial zero one.
Thus for any $b\neq 0,1$ the state $\sigma_b$
violates our criterion from the theorem
(statement (ii)) being then inseparable together
with its ``dual'' counterpart $\sigma_b^{T_2}$.
Here again the limit cases correspond to separable states.
Namely we have:
\begin{equation}
\sigma_0={2 \over 8}P_{\Psi_i} + {1 \over 8}(P_{e_1\otimes e_4}+
P_{e_4\otimes e_1})={1 \over 8}\int_0^{2\pi}
P_{\psi(\phi)}\otimes Q_{\Psi(\phi)}{d\phi \over 2\pi}
\label{in}
\end{equation}
where
$\psi(\phi)=1/\sqrt{2}(1,e^{i\phi})$
and
$\Psi(\phi)=1/2(1,e^{-i\phi},e^{-i2\phi},e^{-i3\phi})$.
Putting b=1 we obtain again the separable state state $P_{\Psi_1}$.

Thus we have provided the families of inseparable states
with positive partial transposition. It is natural to
ask how they are related to the necessary and sufficient
separability condition given in terms of positive maps
( \cite{my} , see Introduction). Clearly it follows that,
in the presence of inseparability of states $\varrho_a$, $\sigma_b$,
there must exist positive maps
$\Lambda_a : {\cal B}({\cal C}^3) \rightarrow {\cal B}({\cal C}^3)$
and
$\Lambda_b : {\cal B}({\cal C}^4) \rightarrow {\cal B}({\cal C}^2)$
such that the operators $I\otimes \Lambda_a\varrho_a$ and
$I\otimes \Lambda_b\sigma_b$ are not positive i.e.
each of
them possesses at least one negative eigenvalue.
It is easy to see that the maps $\Lambda_a$, $\Lambda_b$
can not be of the form
\begin{equation}
\Lambda=\Lambda^{CP}_1+\Lambda^{CP}_2T,
\label{cp}
\end{equation}
where $\Lambda_i^{CP}$ are completely positive maps and T is
a transposition \cite{my}. However the nature of the positive
maps which are not of the form (\ref{cp}) is not known yet
and finding the maps  $\Lambda_a$, $\Lambda_b$ revealing the inseparability
of the states $\varrho_a$, $\sigma_b$ may be difficult.
\section{Conclusion}
We have pointed out that any separable state
can be written as a convex combination of
only $N$ pure product states ($N\leq (dim{\cal H})^2$).
We have provided a new necessary condition
for separability of quantum states in terms of range
of density matrices.
For any separable state it must be possible to
span its range by system of such product vectors
that their counterparts obtained by partial complex conjugation
span range of partial transposition of the state.
It is interesting to see that the above criterion sometimes
does not reveal inseparability
in cases where the partial transposition one works (it can be see for the case
Werner \cite{Werner} $2\times 2$ states) but
it happens to be efficient where the latter fails.
Thus, both the criteria are, in general,  {\it independent}
for mixed states, although one can easily verify
(via Schmidt decomposition) their equivalence for pure states.
One could suppose that, taken jointly, they can constitute
the new necessary and sufficient condition of separability in
higher dimensions. However it is not the case \cite{Peres1}:
one can take the states
$(1-\epsilon)\varrho_a +\epsilon I/9$, $a\neq 0,1$, 
($(1-\epsilon)\sigma_b +\epsilon I/4$, $b\neq 0,1$).
Those states will obviously satisfy the partial
transposition criterion and they would also 
satisfy the present one as the latter is useful 
only for the states with range essentially less then ${\cal H}$.
Now, as the set of separable states
is closed, taking sufficiently small $\epsilon$ one can ensure that
the new states remain inseparable. And this fact is present despite
they satisfy both mentioned criteria.

The present criterion allowed us to provide examples of
states of a new kind, where the entanglement is
masked in a specific way by a classical admixture.
In this context an interesting problem
arises whether it is possible to distill such an entanglement
using local operations and classical communication.

The author is indebted to A. Sanpera for drawing his
attention to the problem and for useful discussions.
He also thanks T. Figiel and J. Popko for consultations on convex sets theory and
R. Alicki, R. Horodecki and M. Horodecki for helpful comments.
Special thanks are due to A. Peres for remarks
leading to significant improvement of the letter.
This work is supported in part by Polish Committee
for Scientific Research, Contract No. 2 P03B 024 12.

\end{document}